\documentstyle[prb,aps,twocolumn,amssymb,amsfonts,epsfig]{revtex}

\begin{document}


\title{Spin fluctuations, electron-phonon coupling and superconductivity in near-magnetic elementary metals;
 Fe,Co,Ni and Pd.}

\author
{T. Jarlborg}

\address{D\'epartement de Physique de la Mati\`ere Condens\'ee,
Universit\'e de Gen\`eve, 24 Quai Ernest Ansermet, CH-1211 Gen\`eve 4,
Switzerland} 

\date{\today}
\maketitle


\begin{abstract}

An investigation of possibilities for superconductivity mediated by spin fluctuations in some elementary metals
is motivated by the recent discovery of superconductivity in the hcp 
high-pressure phase of iron.
 The electronic structure, the electron-phonon
coupling ($\lambda_{ph}$) and the coupling due to spin-fluctuations ($\lambda_{sf}$) are calculated 
for different phases and different volumes for four elementary metals. The results show that such
possibilities are best for systems near, but on the non-magnetic side of, a magnetic instability. Fcc Ni,
which show stable magnetism over a wide pressure range, is not interesting in this respect. Ferro-
and antiferro-magnetic fluctuations in hcp Fe contribute to a relatively strong coupling in the
pressure range where
superconductivity is observed. The absence of fluctuations at large q-vectors makes fcc Pd only moderately
interesting despite its large exchange enhancement for q=0. Fcc Co at high pressure ($\sim$ 0.5 Mbar) behaves as an
improved version of Pd, where the fluctuations extend to larger q. The estimations of T$_C$, which reproduce
the experimental situation in Fe quite well, suggest a measurable T$_C$ for the high-pressure phase of
fcc Co,
while the estimate is lower for the ambient-pressure phase of fcc Pd. 

\vspace{0.3cm}

{\it E-mail address:} thomas.jarlborg@physics.unige.ch

\end{abstract}

\pacs{74.70.Ad,74.20.Mn}

\section{Introduction.}
Superconductivity has recently been reported in the hcp high-pressure phase of iron \cite{shim,jac}.
The maximum of the superconducting transition temperature T$_C$, about 2 K, is in the middle
of a pressure range of 150-300 kBar where superconductivity has been observed. 
Superconductivity is traditionally
explained from electron-phonon coupling. This mechanism is sensitive to magnetic interactions and
it would not be expected to find superconductivity
in magnetic or almost magnetic materials. Alternatively, superconducting pairing can be mediated by
spin fluctuations, and it might be that hcp Fe should be explained by this type of mechanism.
 since anti-ferromagnetic (AFM) instabilities are likely
in the interesting pressure region. Further, some magnetic or nearly magnetic compounds have also been found to be
be superconducting \cite{saxe,pfle}, although the T$_C$'s are not very high. Both the electron-phonon
coupling and ferromagnetic (FM) instabilities require a large electronic density-of-states,
and the question is if superconductivity in such cases is the standard one
caused electron-phonon coupling where T$_C$ is limited by spin fluctuations,
or if it is mediated by spin fluctuations in equal spin pairing (ESP). The fact that the superconductivity
is found only in very pure samples is an indication that mechanism is the latter one, since impurity
scattering is expected to interfere with ESP in a similar manner as magnetic impurities interfere with
pairs which are hold together by electron-phonon coupling. 
In addition, theory indicates the pressure dependence of T$_C$ 
in Fe is inconsistent
with the mechanism of electron-phonon coupling \cite{mazi,pla}. The recent discoveries of magnetic
superconductors led to a 
renewed interest in theories of calculating the coupling strength, $\lambda_{sf}$, for spin fluctuations 
and how such fluctuations can mediate superconductivity \cite{pla,scr,don,tj86,fay,enz,san,mazin:97}. 

Superconductivity in some borocarbides, containing magnetic rare-earth elements is also surprising
and interesting \cite{can}, but the situation is probably quite different from Fe and the compounds mentioned above,
since the magnetic f-electron of the rare-earth atom is very localized and electronically isolated
from the itinerant valence electrons. High-T$_C$ copper oxides might be superconductors because of
AFM fluctuations \cite{pine}, but stripe-like spin waves may interact with phonons \cite{tj01}, and the
mechanism not yet clear. 

This work consider itinerant electrons in pure metals
where magnetic fluctuations and superconductivity must coexist. Near-magnetic pure elements are
particularly interesting in studies of superconductivity and magnetic fluctuations, both
experimentally and theoretically. Non-compound elements can be made very pure without intrinsic defects caused by 
non-stoichiometry or vacancies in sub-lattices of compounds. This
allows for ESP to develop without much interference from impurity scattering. Theoretically it is
also preferable to consider pure metals over compounds. The electronic structure is simpler and
especially for studies of extended AFM fluctuations by the method of frozen spin waves, it
avoids the complication of having several types of sites where the local magnetic response can be different.
In this work we study four metals, Fe, Co, Ni and Pd, which all are known to be ferromagnets
or strongly enhanced paramagnets. Different phases are considered, and total energy calculations
are used to determine the stability of hcp, fcc and bcc phases as function of pressure.
A previous report considered FM and AFM spin fluctuations separately in hcp Fe, and it was concluded
that the combined effect could explain the observed T$_C$ \cite{pla}. This study is here extended
to include the other three metals and where
a weighting of different fluctuations are made, similar to what is made for phonons.

\section{Theory.}

The electronic structures are calculated using the linear muffin-tin orbital method 
in the local spin density approximation (LSDA) \cite{lmto,lda} and the 
generalized gradient approximation (GGA \cite{gga}) scheme, which includes non-local corrections to the
potential. The basis set includes states up to $\ell$=3.
The c/a-ratio in the hcp structures is fixed to 1.59 for all volumes. 
This is close to the calculated value within a wide pressure range for Fe \cite{stix}, and
a few tests with other ratios show that this choice is good also for Co.
The number of k-points vary with the structure. For fcc and bcc it 505 and 506 points in the irreducible part of the
Brillouin Zone (BZ), for hcp 252, 
and for the supercells with long wavelength AFM configurations we use the same density of points. 
The number of points is close to 500/n where n is the number of atoms per cell.
The use of GGA instead of LSDA potential is known to be crucial for a correct description of the fcc vs. bcc
stability of Fe, but other properties are not very different in the two types of density functional 
potentials \cite{moro}. A previous report on Fe used mainly LSDA for the calculations of $\lambda_{sf}$ \cite{pla}.
Here, these results are complemented with calculations using GGA for the most interesting configurations
in hcp iron. The magnetic coupling constants are similar (within $\sim$ 5 percent) to those calculated 
in the LSDA. As in ref. \cite{stix} it is found that two AFM configurations compete
with the NM state in hcp iron at some pressures. 
The first (AFM-I) has opposite polarization on each alternate z-plane layers,
while the second (AFM-II) has opposite polarization on each layer perpendicular to the x-axis. 
The results for 
Ni, Co and Pd are new, and only results from GGA calculations are reported.

The electron-phonon coupling parameter $\lambda_{ep}$ is calculated from the electronic structure; 
\begin{equation}
\label{eq:lambda}
\lambda_{ep} = N V^2 / M \omega^2 = N V^2 / K
\end{equation} 
where $N$ is the density of states at $E_F$, $M$ the atomic mass, $\omega$ a phonon frequency and $K$ a force constant.
The latter is related to the change of total energy $E$ due to an atomic displacement $r$, $ K = d^2E/dr^2$.
The Hopfield parameter $N V^2$ contains
$V$, the Fermi surface (FS) average of the matrix element $< \psi_k (dv(r)/dr) \psi_k >$, 
the first order change in energy due to the displacement,
where $\psi_k$ is the wave function at the k-point $k$, and $v(r)$ is the potential.
For elementary metals it is a good approximation to calculate $V$ within the rigid muffin-tin approximation 
RMTA \cite{gg}, 
which in the case of LMTO band results is slightly modified to the rigid Wigner-Seitz (WS) approximation \cite{daco}. 
The alternative to calculate $V$ from the change in band energies, $\varepsilon_k$, when the structure
is deformed as in a 'frozen' phonon, demands a larger computational effort since several large supercell
calculations are needed to map out representative phonons. 
The force constant is fitted to the experimental value of the Debye temperature $\Theta_D$ at zero pressure,
and then scaled to other volumes by the calculated bulk modulus \cite{pict}. This scaling is assumed to
be valid also between the hcp and bcc phases for Fe and between the hcp and fcc phases of Co.
The calculated values of $\lambda_{ep}$ at equilibrium volumes 
agree well with independent results for the elements considered 
here \cite{papa,jp}. This is despite the use of Wigner-Seitz cells instead of non-overlaping Muffin-Tin spheres in the
evaluation of the matrix elements.

Interesting spin fluctuations occur near magnetic instabilities when only a small energy
is needed to excite a magnetic wave in the material. The energy is zero at the magnetic
instability and it can be very low close to the instability, lower than the energy of typical phonon energies. 
It is also noted that spin fluctuations can be the result of a divergence of the
generalized susceptibility, $\chi(q,\omega)$, implying that Fermi surface nesting for critical
values of $q$ is partly responsible for the divergence. This is also the case for phonon
softening when some $q$-modes become unstable. Phonons with $q$ outside the BZ have no sense,
since the wave length cannot be shorter than two interatomic distances. Spin- or charge- density waves
with wavelengths shorter than the size of an atom can be imagined, 
but if they were important they should probably exist already on the level
of a single atom, before the scattering within several sites in a lattice can set up a FS and nesting.
Since the structure always seems to be important for the properties of a material, we assume that 
fluctuations 
corresponding to $q$-vectors outside the BZ are not important. Thus, we consider only "soft"
spin fluctuations with 
energies and wavelengths which are similar as for phonons. 
If adiabatic conditions and the Migdal theorem are valid for
phonons they should also be valid for soft spin fluctuations. 

As was pointed out in a previous work \cite{pla}, the analogies between soft spin-wave excitations
and phonons can be utilized. A spin wave with a moment $m$ and a phonon with lattice-displacements "$r$" 
both generate perturbations of the potential, although the perturbations are of different kinds. 
Magnetic moments and the spin-polarized perturbation of the potential have spherical symmetry within each
site, except for very short wave lengths of the spin waves. 
The matrix elements $V$ are mainly monopolar, which couple 
electronic states of equal $\ell$. 
Atomic displacements generate mainly an asymmetric (and non-polarized) perturbation
of the potential at each site. Ionic compounds, where the displacements are associated with large Madelung
shifts, can have some monopolar coupling \cite{hpa}. But here, in neutral pure elements, the matrix element $V$ is
dominantly dipolar with the selection rule $\ell - \ell ' = \pm 1$.   
In addition there is a technical difference of how the frozen phonon and 
spin wave calculations are made.
By applying a magnetic field $\xi$ one obtains a magnetic moment $m$ in frozen spin wave configurations. However,
the displacements in frozen phonon calculations are imposed instead of being a result of an applied force.
 This difference would disappear if we did so-called 
"fixed-spin-moment" (FSM) calculations
instead of applying the magnetic field. The total energy in such case is the free energy for the configuration
with the fixed moment, while with imposed field the free energy is the total energy plus the term field times
moment. In practical calculations we prefer the method with applied fields which is simple to use both for FM
and AFM cases.  

The calculation of mass enhancement due to FM or AFM spin fluctuations, $\lambda_{sf}$, is analogous
to that for phonons, but with "$r$" replaced by "$m$".

\begin{equation}
\label{eq:lam}
\lambda_{sf} = N V^2 / K = N <\psi_k (dv(r)/dm) \psi_k>^2 / (d^2E/dm^2)
\end{equation}

Calculations for spin wave excitations with general q-vectors require supercells, 
as in the case of frozen phonon calculations.
The calculations are made for two different configurations, one with an applied field
and one without, to give two band structures and two magnetizations. If the changes of the
free energy $(E)$ are harmonic, i.e. quadratic functions of the induced magnetic moment, 
we have $E_{\xi}=E_0 + J m^2$
where $m$ is the moment (induced by the field $\xi$) per atom.  
 The constant $J = \frac{1}{2}d^2E/dm^2 = (E_{\xi} - E_0)/m^2$ is calculated from spin polarized results, 
 where the applied fields range from 1 to 10 mRy. The local Stoner enhancement, $S$, is
defined as $(\epsilon_{\xi} - \epsilon_0)/\xi$, where $\epsilon_{\xi}$ is the increase of the exchange splitting 
at an atom (obtained from the logarithmic derivative
of the d-band) induced by the field $\xi$. (K-dependent Stoner enhancements can be obtained by taking the
exchange splitting from the band energies at each k-point.) The $S$-values are generally smaller for large $\xi$, 
indicating some aversion against large moments and non-harmonic variations of $J$ as function of field.

The monopolar matrix element $ <\psi_k dv \psi_{k'}>$ can be calculated directly from the change in the spin part
of the potentials between the two cases, with and without field. This permits calculation of the anisotropy
of the coupling, $\lambda(k,k')$, which is needed in Eliashberg theory to determine if the superconductivity
is s-, p-wave or of some higher symmetry. In this work we do not determine the anisotropy, but
only the FS average of $\lambda(k,k)$. In this case it is
convenient 
to determine the matrix element directly from the change in band energies, simultaneously to the 'frozen spin wave'
determination of the change in total energy. Hence,
$ <\psi_k dv \psi_k>^2 = (\varepsilon_k^m-\varepsilon_k^0)^2$, where $\varepsilon_k^m$
is the band energy at point $k$ for the configuration belonging to $m$.  
In the harmonic limit we obtain finally;
\begin{equation}
\label{eq:lam1}
\lambda_{sf} = N < (\varepsilon_k^m-\varepsilon_k^0)^2 >_{FS} /2(E_m-E_0)
\end{equation}
Here $< ~ >_{FS}$ means a FS average.
The matrix element
is more efficient for FM than for AFM fluctuations, because a FM field applied in a system
where one band is dominant at $E_F$, will split all bands about equally. The same difference 
$(\varepsilon_k^m-\varepsilon_k^0)$ will appear almost everywhere over the FS. For induced AFM configurations
it is more difficult to visualize these differences. Positive and negative differences are likely, which means
that small differences should exist at some sections of the FS and make the averaged matrix element smaller.

The energy of the FM and AFM fluctuations, $\hbar \omega_{sf}$, is estimated from a Heisenberg model, where the 
parameter $J$, the exchange integral, is the maximum amplitude of the spin wave \cite{kittel}.
For ferromagnetic fluctuations it has been proposed that $\hbar \omega_{sf} = 1/(4NS)$ \cite{mazin:97}.
The corresponding property in the present approach is $J/4$.

In order to make the calculations of superconductivity mediated by spin fluctuations analogous to
that mediated by phonons, one needs to include a number of representative $q$-vectors. We consider
FM fluctuations ($q$=0) and a few AFM fluctuations with $q$ at the zone boundary. For fcc Co we
calculate intermediate $q$ by use of supercells with up to 8 atoms. The results from the intermediate
$q$ vectors show no surprises, but the coupling and $J$ are between the values for large and zero $q$.
This indicates that no drastic effects from the FS are seen, but it can not be excluded that effects from
nesting at particular $q$-vectors can be missed by the method of supercell calculations. It is as for
frozen phonon calculations, where one assumes a smooth variation between the values calculated at
a few $q$-points. If a special mode-softening
is known initially it can be studied carefully. 

The strong coupling parametrization of the
Eliashberg theory for electron-phonon coupling by McMillan \cite{mcm} contains the weighted
averages of the coupling $\lambda$ and the frequency $\omega_{log}$. 
\begin{equation}
\label{eq:mcm1}
\lambda = 2 \int \alpha^2 F(\omega) \frac{d\omega}{\omega}
\end{equation}
\begin{equation}
\omega_{log} = exp(\frac{2}{\lambda} \int \frac{ln \omega}{\omega} \alpha^2 F(\omega) d\omega)
\end{equation}

With the few number of spin waves we replace the integration by a summation
over $q$-points "$i$" with relative weights $w_i$. The density-of-state contribution $\alpha^2 F(\omega_i)$
from mode $i$ is contained in the value $\lambda_i$, which is defined
$2 \Delta \omega_i \alpha^2 F(\omega_i)/ \omega_i$
for a single mode.

\begin{equation}
\label{eq:mcm2}
\lambda = \sum_i w_i \lambda_i
\end{equation}
\begin{equation}
\label{eq:mcm2b}
\omega_{log} = exp(\frac{1}{\lambda} \sum_i ln \omega_i  w_i \lambda_i)
\end{equation}

Superconductivity based on electron-phonon interaction is easily destroyed by magnetic impurities
and magnetic fluctuations. The spin-splitting of the potential leads to
an attraction between electrons of equal spin and compete with 
the attractive interaction between Cooper pairs. Effects from impurities are not included, but
for the effect from spin fluctuations we use the
Daams et al modification \cite{daams} of the McMillan formula due to $\lambda_{sf}$, 
so that $\lambda_{eff}=\lambda_{ep}/(1+\lambda_{sf})$ will replace $\lambda_{ep}$. 
\begin{equation}
\label{eq:tcep} 
k_{B}T_{C} = (\frac{\hbar \omega_{log}}{1.2}) exp(\frac{-1.04(1+\lambda_{eff})}
{(\lambda_{eff}-\mu^*(1+0.62\lambda_{eff}))})
\end{equation}
In addition, with $\mu^*$=0,
the expression becomes formally very similar to the expression for T$_C$ by paramagnon coupling,
see below.

For superconductivity mediated by spin fluctuations the situation is reversed, so that non-magnetic
impurities and strong electron-phonon coupling reduce T$_C$. This is because the pairing is
between electrons of equal spins, and scattering with impurities or phonons will reduce the 
spin-splitting of the local potential. 
For estimating the superconducting transition temperature we use a simplified version
of the weak coupling BCS-like formula
for paramagnon coupling  \cite{bcs,fay}.
\begin{equation}
\label{eq:tcsp}
k_{B}T_{C} = ({\hbar \omega_{sf}}/{1.2}) exp ({-(1+\lambda_{ep}+\lambda_{sf}^0)/ \lambda_{sf}^1})
\end{equation}
The calculated enhancement parameter due to spin fluctuations is mainly 
considered to be of p-character, $\lambda_{sf}^1$, 
although we have not investigated the $(k,k')$-dependence over the FS.  
The s-wave fraction of the total $\lambda_{sf}$,
$\lambda_{sf}^0$, is expected to be small, but it can be more important at
larger q, when the perturbing potential has stronger dipolar character, as in the 
case of atomic displacements in 
electron-phonon coupling. The s-part should also
contain a contribution from transverse fluctuations \cite{fay}. Two extreme values of $\lambda_{sf}^0$
are assumed, either that $\lambda_{sf}^0$=$\lambda_{sf}^1$ or that $\lambda_{sf}^0$=0, in order
to obtain the range of extreme values on T$_C$.
The second choice turns out to give a $T_C$ of a
reasonable amplitude compared to what is measured for hcp Fe, while the T$_C$ is about half as large
if the s-part is assumed to be equal to the p-part.  It is obvious that with the various approximations
it is not possible to give precise values for $T_C$, but the values should be taken as a possible
range of T$_C$ in the different materials. The result for hcp Fe serves as a calibration; since the T$_C$
and its pressure variation agree quite well with experiment, there is some hope that the same
method can be used as a guidance for the other materials, especially for fcc Co where no experimental
data are known.

\section{Results.}

The exchange enhancement is defined as the increase of the exchange splitting of the d-band
divided with the energy of the applied magnetic field \cite{pd}. This definition is
natural for FM cases and it leads to the usual Stoner enhancement for paramagnets. 
We also define a local exchange enhancement for AFM configurations, when the field 
is applied differently on different sites. The results show that these exchange enhancements
are largest when the material is close to, but on the non-magnetic side of, 
a magnetic (FM or AFM) instability. As the volume is decreasing in a magnetic system one
will find a critical volume where the (local) moments disappear at a certain volume. The local
exchange enhancement will be infinite at the critical volume and still large for decreasing
volumes. But within the region where the volumes are larger than the critical volume, 
the exchange enhancement drops rather quickly. It is as if the minimum of the total
energy as function of the moment in a magnetic system is deeper than in a non-magnetic one,
and the additional fields will not have a large influence on the moment. In other words,  
non-magnetic materials are magnetically "softer" than the materials which
already have a magnetic ground state. 
A consequence of this is that $\lambda_{sf}$ will behave asymmetrically
as function of pressure around the critical volume, so that it will be lower in magnetic
region than in non-magnetic one. This is at least the results from the calculations
for several configurations in these pure elements.
One may add that we have not considered more than the three most common structures.
The calculations can not exclude that other structures may appear at high pressure.

\subsection{Nickel.}

 The calculations for fcc-, bcc- and hcp-structures of Ni
show that the FM state is very persistent. This is for pressures varying from zero up to
more than 3 Mbars, or when the volume is compressed to 45 $(a.u.)^3$/atom from the equilibrium volume
of 73 $(a.u.)^3$/atom. 
Only the bcc structure has a vanishing moment for volumes below about 62 $(a.u.)^3$/atom,
but the total energy is always larger than for magnetic fcc, about 
4 mRy/atom within the whole range of volumes.
The hcp structure is FM with a moment ranging from 0.6 to 0.49 $\mu_B$/atom in this
range of volumes. This is very similar to fcc, where the range of $m$ is from 0.62 to 0.48 $\mu_B$/atom.
However, the total energy of the hcp phase is 3-5 mRy/atom larger than in fcc. 
The results suggest that NM bcc should be stable over FM hcp at large pressure, but FM fcc
is the most stable structure everywhere.

These facts, the stability of the FM fcc phase, and the low values of exchange enhancements 
and $\lambda_{sf}$ for magnetic
phases, make together that Ni seems not interesting as a candidate for superconductivity
mediated by spin fluctuations. The NM bcc phase could have large $\lambda_{sf}$ near the
magnetic instability for a pressure of the order 0.5 Mbar, but since it is not the most
stable phase we refrain from making detailed calculations. 

\subsection{Iron.}

Several results for Fe in a previous publication \cite{pla} are complemented
by adding a few GGA results. 
The FM bcc phase is stable for volumes larger than 72 $(a.u.)^3$/atom, while NM hcp is
stable for lower volumes, see fig. 1. Two AFM configurations are stable for hcp. The AFM-I configuration
(when the moments alternate along the z-axis)
is stable only at large volumes, while the AFM-II configuration (alternation along the x-axis)
is stabilized just below
the critical volume where bcc is stable. This is in agreement with independent calculations, where
interpretation of Raman shifts lends support to the existence of AFM order \cite{stein}.
However, the moment is small and the total energy
is almost the same for the NM hcp structure for volumes smaller than 72 $(a.u.)^3$/atom \cite{pla}.
The calculation of $\lambda_{ep}$ is done in the rigid WS approximation with
the phonon denominator scaled by Debye temperature and the P-dependence of the bulk modulus.
The $\lambda_{ep}$ for the bcc structure agree with previous RMTA calculations \cite{papa,jp}. The
values for the hcp phase are somewhat smaller than in ref. \cite{mazi}, with a similar relative 
decrease as function of pressure (P).
This is mainly because of the difference in force constant $K$, while the matrix elements for the electron
phonon coupling are similar. 

The calculations of $\lambda_{sf}$ and T$_C$ are done following the procedures described above
using the previous results from FM and the two AFM configurations \cite{pla}. The weights in 
the summations for $\lambda_{sf}$ and $\omega_{log}$ (1/5, 1/5 and 3/5 for FM, AFM-I and
AFM-II, respectively) are obtained from the q-values of the spin configurations.  The results are 
given in Table 1. The variation of $\lambda_{ep}$ is continuous between the largest and
smallest volume. This is true also for $\lambda_{sf}$ and $\omega$ for
FM fluctuations. The FM instability occurs well beyond the range of volumes
considered here where the Stoner enhancement never is larger than 3.2. The corresponding values
$\lambda_{sf}$ (and $\omega$) for the two AFM configurations show pronounced
peaks (dips) near the critical volumes for the magnetic instabilities. The local Stoner factors
diverge at these volumes.  These instabilities are reflected in the averaged values of
$\lambda_{sf}$ and $\omega_{log}$, so that together with the contribution from FM fluctuations
there are conditions for superconductivity near two volumes, see fig. 2. The calculated T$_C$
agrees well with experiment when $\lambda_{sf}^0$ is assumed to be zero, with a maximum
of about 1.2 K just below the volume at which the hcp phase becomes stable and with T$_C$ approaching
zero at larger pressures. A second peak in $T_C$ is seen at larger volume, where the AFM-I
becomes unstable. However, this can not be tested experimentally since the bcc structure is stable at    
such volumes. Also hypothetically, it is possible that strong FM fluctuations may give superconductivity
at even larger volumes. As a comparison we also calculate the T$_C$ based on electron-phonon 
coupling with reduction due to $\lambda_{sf}$, using the formula of Daams et al \cite{daams}.
With $\mu^*$=0, T$_C$ is about 1.0 K just after the bcc-hcp transition at the volume 70 $(a.u.)^3$/atom.
This value is in reasonable agreement with experiment, but the reduction with pressure is
too slow in comparison with experiment. At 58 $(a.u.)^3$/atom T$_C$ is still 0.3 K. This
corresponds to a rather smooth reduction of T$_C$ from 1.0 to 0.3 K within a pressure range of
 $\sim$ 700 kBar, compared to a confinement of T$_C$ within 200-300 kBar from spin fluctuations
 and within $\sim$ 150 kBar experimentally \cite{shim}. Thus, from the pressure dependence of T$_C$ it
 can be concluded that superconductivity is more likely to be mediated by spin fluctuations. 
A similar conclusion was reached by Mazin et al \cite{mazi}.

The good agreement with experiment for the amplitude of T$_C$ could be accidental. Many approximations
are used and the statistics for the q-average is poor. On the other hand, the relative variations
with pressure should be more reliable, since the procedure for calculating T$_C$ always is the same.
This motivates a study of Co and Pd with the same procedure in order to verify if Co could be
superconducting at high pressure and if fcc Pd, with its large exchange enhancement, will
incorrectly appear as a superconductor by this theory.

\subsection{Cobalt.}

The stable phase of Co at zero pressure is FM hcp with a moment of 1.59 $\mu_B$/atom when volume
is 73.1 $(a.u.)^3$/atom. The total energy of the NM hcp state is 15.5 mRy/atom higher at the same
volume, which is close to the experimental volume at equilibrium. Figure 3 shows the total energies
from FM and NM GGA calculations of the three structures, and the FM moments are shown in fig. 4.
The bcc structure is magnetic within a wide range of pressures, but its total energy is
considerably above those of fcc and hcp, and the bcc phase is not of interest here. 
It is seen in fig. 3 that the fcc structure becomes stable (over hcp)
at a volume near 58 $(a.u.)^3$/atom. The pressure at this volume is about 0.8 and 0.7 Mbar
for hcp and fcc, respectively. 
The transition volume is below the volume (about 60 $(a.u.)^3$/atom) where fcc becomes NM.
Therefore, from the experience with Fe it is expected that the Stoner enhancement and 
$\lambda_{sf}$ for FM fluctuations become very large at volumes just below the FM instability.
In spin polarized calculations near the critical volume for the FM instability
it is possible to find metastable low- and high-moment states
with almost degenerate total energies. This leads to large non-linear Stoner enhancements
so that S are large at large field. The S-values vary from 4 at the volume 54 $(a.u.)^3$/atom
when the applied field is 1 mRy to 28 at 59 $(a.u.)^3$/atom for the field of 2 mRy.
As was discussed earlier, it is generally easier to find large $\lambda_{sf}$  for FM than for AFM
fluctuations \cite{pla}, and here we obtain very large values, from 0.9 to 2.4 within this range
of volume and field. This is much larger than for FM fluctuations in hcp Fe, where the FM
instability is outside the considered range of volumes. The AFM instabilities are within
the range volumes, but the corresponding $\lambda_{sf}$ is rather modest,
since the matrix element for AFM fluctuations is not so efficient. 

The FM Stoner factors on the magnetic side of the transition are much smaller, about 1.7, and typical
$\lambda_{sf}$ is of the order 0.1. This agree with the findings for AFM fluctuations in Fe, and
FM fluctuations in hcp Co. The latter is magnetically ordered within the entire range of volume
where it is the stable phase, and it is not promising for superconductivity.

A few AFM configurations are made for two volumes of fcc Co just below the volume where FM disappears, at
56 and 59 $(a.u.)^3$/atom. The applied field is used to impose AFM modulations in the x-direction
for different wave lengths, corresponding to 2, 4 and 8 atoms in supercell calculations. (The field
amplitude, 1 or 2 mRy, is the same on each site so that the wave is step-like rather than sinus-like.)
The local Stoner enhancements are lower than for the FM case at equivalent volumes. For the
shortest AFM wave $S$ is 2.3 compared to 28 for the FM configuration. For the longest wave
there are 4 layers of atoms where the field is of the same sign. Thus, the two in the middle 
are surrounded by sites of the same polarization and $S$ is 10 within this region of "local"
FM order. But the two atoms at the edge of the 4-layer atoms, have one neighbor with reversed
polarization, and $S$ is lower, about 6, for these sites. The coupling $\lambda_{sf}$ is
relatively small for all of the AFM configurations. It varies from 0.26 to 0.09 and to 0.03 for the
longest, the intermediate and the shortest wave, when the volume is 59 $(a.u.)^3$/atom. These values are reduced further
to 0.07, 0.06 and 0.03, when the volume is 56 $(a.u.)^3$/atom. 

Some values the electron-phonon coupling were calculated for the hcp phase. 
 At large volume this phase is FM and the coupling
is dominated by the minority band which has the largest DOS. At the equilibrium volume $\lambda_{ep}$
is 0.19 for the minority band and 0.05 for the majority band, while at 54 $(a.u.)^3$/atom,
when the magnetism is strongly reduced, the amplitudes
are reversed, 0.25 and 0.20. This is because of a large increase of the majority DOS.
 A $\lambda_{ep}$ of this order
should make superconductivity possible if spin fluctuations are absent. However, hcp Co is 
still magnetic at this volume and the fact
that fcc is the stable phase at this volume makes it not very interesting to study superconductivity
based on electron-phonon coupling. Instead, we calculate $\lambda_{ep}$ for the NM fcc phase in
order to investigate whether it can give superconductivity mediated by phonons or if it
will prevent superconductivity mediated by spin fluctuations. The calculated values of $\lambda_{ep}$
are rather stable as function of volume, about 0.41. This is for a $\Theta_D$ of 550 K at the
volume 56 $(a.u.)^3$/atom. The calculation of T$_C$ is made as for Fe. If the reducing effect
from the (q-weighted) $\lambda_{sf}$ is included one obtains 0.1 K at 56 $(a.u.)^3$/atom
and 0.01 at 59 $(a.u.)^3$/atom. Without influence of spin fluctuations we obtain 0.56 and 0.3 K.

The estimation of T$_C$ from spin fluctuations are done as in the case of Fe, using
the q-weighted results in table 2, with the reducing effect
from electron-phonon interaction and with two assumptions for $\ell=0$ part of $\lambda_{sf}$
(either 0 or equal to the $\ell=1$ part). The results are 0.2 and 0.06 K, respectively, at the
smallest of the two volumes, while 1.1 and 0.4 K, respectively, at the large volume.

There are many approximations and uncertainties behind these values, but since the
method is identical to the method used for Fe, where the experimental amplitude is fairly well
reproduced or even underestimated, there is some hope that superconductivity could be observed in Co
just after the hcp-to-fcc phase transition. The calculations predict the transition pressure 
to be very high, of the order 700 kBar, but as seen in fig 3, a small error in the relative
position between the total energy curve will shift the transition volume (and pressure) a lot.
If the critical pressure in reality were even higher, it would be less likely to expect
superconductivity mediated by spin fluctuations, because the material would be
further away from the instability point when the fcc structure becomes NM. The critical
volume for the FM to NM transition is not as sensitive because if does not depend on the
relative differences between the total energy of two structures. However, the magnetic
transition for the fcc phase is diffuse and extends over some interval of volumes, where
non-linear effects and metastable states are found. Therefore, if the hcp to fcc transition turns out
to happen at larger volume, there is some margin for having larger values of $\lambda_{sf}$
and T$_C$, than what has been calculated here. This is since the volume for the onset of
FM in the fcc structure appears to around 60 $(a.u.)^3$/atom, while the largest volume
for the calculation of $\lambda_{sf}$ is 59 $(a.u.)^3$/atom. All partial values of
the enhancements and $\lambda_{sf}$, FM or AFM, tend to increase when the system
is closer to the critical volume, but non-linear effects make the
supercell calculations more difficult.

Some differences with Fe can be noted. The structures are different and it is difficult
to say if this somehow will annihilate the advantages coming from the fact that the
method of calculation is as identical as possible for the two systems. The relative weights
between FM and AFM fluctuations are also different. Fe is far from the FM instability,
but there two stable AFM configurations are close and fluctuations from them contribute
to the final $\lambda_{sf}$. No AFM instability is found in these calculations for fcc Co.
It is the FM case which dominates, with a Stoner factor which at the largest volumes
becomes considerably larger than in Pd. Since Pd at ambient pressure also is of fcc structure,
has a large exchange enhancement, but has not been found superconducting, one could test
the same method for Pd.

\subsection{Palladium.}

For Pd we consider only the fcc phase at the experimental lattice constant when the volume
is 98.5 $(a.u.)^3$/atom. The electon-phonon coupling is calculated to be 0.36 for
a $\Theta_D$ of 274 K. The Stoner
enhancement for FM configuration is 5.7, in agreement with earlier estimations using LSDA \cite{pd}.
This enhancement is smaller than the largest values calculated for Co, which implies
that the hypothetical FM instability in Pd would require a considerable increase in volume.  
Also the local enhancements for the
AFM configurations are smaller than the corresponding values for Co. From the calculations
using 2 and 4 atoms/cell $S$ is 1.7 and 1.9, and for the two inequivalent sites in the 8
atom cell the values are 2.4 and 3.6. The coupling $\lambda_{sf}$ for the three AFM cases
are of the order 0.02 to 0.14 from the shortest to the longest wave, while for the FM
case it is calculated to be between 1.4 and 1.8 depending on the amplitude of the applied field.
Thus, the enhancements in Pd is comparable to the result for Co calculated for the smallest
volume. The q-averaging is done for the equivalent modes as for Co,
and the results are  $\lambda_{sf}$=0.15 and $\omega$=13 mRy.

The estimate for T$_C$ is below 0.01 K for superconductivity mediated by electron-phonon coupling
(but limited by spin fluctuations), and between 0.08 and 0.03 K for the two assumptions 
of $\lambda_{sf}^0$ in the calculation of T$_C$ from ESP (and limited by electron-phonon
coupling). The latter values are smaller than the results for two volumes of fcc Co, and T$_C$ is very
low despite the large exchange enhancement of Pd. 
The reason is that the large exchange enhancement is not extending to large q-values,
and the averaged $\lambda_{sf}$ is not large enough for superconductivity. The material does
not support spin excitations where adjacent atoms have different polarization. This
is very different from hcp Fe where opposite polarization on near neighbors even give
stable AFM configurations. If the volume of Pd were $\sim$ 5 percent larger, one could
expect results similar to what is calculated for fcc Co. The reduction of enhancements at
large q is crucial for an understanding of the small T$_C$ or even absence of superconductivity.
A calculation based only on the standard Stoner enhancement (at q=0) would not be correct.
A recent work suggests that spin-orbit coupling can suppress superconductivity in Pt \cite{fay2}.
One may speculate that such an effect should be more evident in Pd than in Co because of the relative
strengths of the spin-orbit coupling.

\section{Conclusion.}

Two possibilities for superconductivity in four exchange enhanced metals have
been investigated by calculations of the coupling constants for electron-phonon
coupling and spin fluctuations. The method for calculating $\lambda_{ep}$ and the corresponding
T$_C$ has been quite successful in many studies during the last decades, while the method
for calculating $\lambda_{sf}$ is new and needed some testing. Both methods include
approximations and one should not rely too much on to the absolute values of T$_C$,
but look more at the relative variations among the different cases and different pressures.
Despite these precautions, it seems clear that superconductivity based electron-phonon coupling is unlikely in Pd,
Co and Ni. This is because of the limitation of T$_C$ coming from spin fluctuations or from
stable magnetic moments, and from the rather modest values of $\lambda_{ep}$. Hcp Fe could
have a reasonable T$_C$, but the agreement with experiment is poor for the pressure dependence
when electron-phonon coupling determines T$_C$.

The results for T$_C$ based on ESP are partly a test of the method, partly a prediction.
The experimental situation in hcp iron is surprisingly well reproduced, both the amplitude
of T$_C$ ($\sim$ 1.2 K) and its variation with pressure. This result indicate that a representative q-point
average of $\lambda_{sf}$ has to be calculated. The calculation for fcc Pd pass the test as well,
since T$_C$ is very low ($\sim$ 0.08 K) despite its large Stoner factor. Again, this is the result only when
several configurations are weighted together. 
The calculations indicate that a NM fcc phase of Co becomes stable at a high pressure of the order
half a Mbar. Here, Co behaves as an improved version of Pd with larger enhancements, but with
similar relative variation as function of q. The improved conditions for superconductivity 
from spin fluctuations give a T$_C$ of the order 1 K.

These calculations do not include considerations of defects.
Since superconductivity due to FM fluctuations is sensitive to impurity scattering \cite{foulk}, it is 
possible that lattice defects, induced in pressure experiments near a structural transition,
can suppress $T_C$. It is not yet clear if experimental conditions can be improved to
make T$_C$ higher in hcp Fe or even in Pd. 
The reduction of $T_C$ compared to $T_C$ itself is proportional to $v_F/(\ell \Delta)$, where 
$\ell$ the impurity scattering length, $\Delta$ the superconducting energy gap and
$v_F$ is the Fermi velocity \cite{fay}.
Knowledge of the q-dependence of the destructive effect from defects on $\lambda_{sf}$ or 
ESP should be useful, since this can make a difference between Fe and Co (and Pd).
The FM fluctuations are relatively important in the latter two metals of fcc structure.
This is in contrast to hcp Fe, where the AFM fluctuations have a large weight.   
Calculations for supercells containing a defect give information about
how much a defect will hamper the development of FM or AFM spin waves.
The real space interpretation of the AFM results for Pd is that it is not easy for neighboring sites to
have different magnetization. If so, it should also be difficult for a FM moment to develop near to
a non-magnetic impurity in Pd, which could make this material more sensitive to impurities than Fe.
In the absence of full calculation with impurities one can also note that
 from the dispersion of spin waves it follows that
fluctuations with intermediate $q$ are mobile with large values of the group velocities. 
They may interfere with defects, while waves with $d\omega / dk \sim 0$
(as for a purely FM fluctuation) can develop easier between defects. 
Co is an intermediate case between Fe and Pd even without considerations of defects, 
still with possibilities for a measurable T$_C$.




\begin{table}
\caption{Volume per atom, partial and q-weighted $\lambda_{sf}$ and $\omega_{sf}$ (in energy units)
for hcp iron. The weights are 0.2 0.2 and 0.6 for FM, AFM-I and AFM-II, respectively. 
The electron-phonon coupling decreases continuously from 0.37 for the largest volume
to 0.14 for the smallest volume.}
\begin{tabular}{ccccccccc}
 V & & $\lambda_{sf}$ & & & & $\omega$ & & $\omega_{log}$    \\
   &    FM & AFM-I & AFM-II & tot & FM & AFM-I & AFM-II & tot  \\
 $(a.u.)^3$ & & & & & mRy & mRy & mRy & mRy    \\
\tableline
79.2 & 0.61 & 0.08 & 0.02 & 0.15 & 3.2 & 6.3 & 25 & 4.1 \\
76.8 & 0.54 & 0.61 & 0.03 & 0.25 & 3.7 & 0.5 & 20 & 1.6  \\
74.5 & 0.43 & 0.30 & 0.04 & 0.17 & 4.0 & 0.9 & 26 & 2.8 \\
70.1 & 0.32 & 0.14 & 0.14 & 0.18 & 5.0 & 2.1 & 3.0 & 3.4 \\
67.0 & 0.28 & 0.09 & 0.26 & 0.23 & 5.5 & 3.0 & 1.4 & 1.8 \\
66.0 & 0.26 & 0.06 & 0.21 & 0.19 & 5.8 & 3.5 & 1.3 & 2.0 \\
62.1 & 0.20 & 0.03 & 0.10 & 0.11 & 7.2 & 4.8 & 2.1 & 3.5 \\
58.5 & 0.15 & 0.02 & 0.06 & 0.07 & 11. & 6.2 & 3.2 & 5.7 \\
\end{tabular}
\end{table}
\begin{table}
\caption{Volume per atom, $\lambda_{ep}$, partial and total q-weighted $\lambda_{sf}$, and total
$\omega_{log}$ (in energy units) for spin fluctuations in
fcc Co at two volumes, and in fcc Pd (last line). 
The weights are 1/16, 3/16, 6/16 and 6/16 for FM, and the 3 different
AFM configurations ordered according to the wave length (2 4 or 8 atoms).
}
\begin{tabular}{cccccccc}
 V & $\lambda_{ep}$ & & & $\lambda_{sf}$ & & & $\omega_{log}$ \\
 $(a.u.)^3$ & & FM & AFM-2& AFM-4 & AFM-8 & tot &  mRy     \\
\tableline
56 & 0.42 & 1.6 & 0.03 & 0.06 & 0.07 & 0.18 & 12.8 \\
59 & 0.41 & 2.4 & 0.03 & 0.10 & 0.26 & 0.27 & 5.6 \\
99 & 0.36 & 1.4 & 0.02 & 0.04 & 0.15 & 0.16 & 13.1 \\
\end{tabular}
\end{table}

\begin{figure}[tb!]

\leavevmode\begin{center}\epsfxsize8.6cm\epsfbox{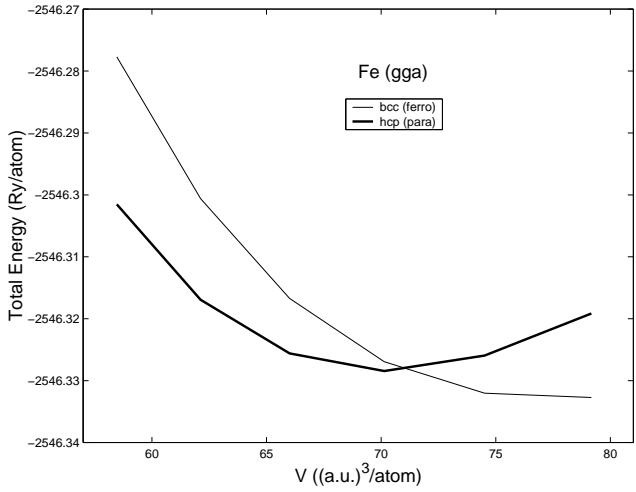}\end{center}
\caption{
 Calculated total energies for paramagnetic hcp (heavy line) and ferromagnetic bcc (thin line)
 iron using the GGA potential.
 }
\end{figure}
\begin{figure}[tb!]

\leavevmode\begin{center}\epsfxsize8.6cm\epsfbox{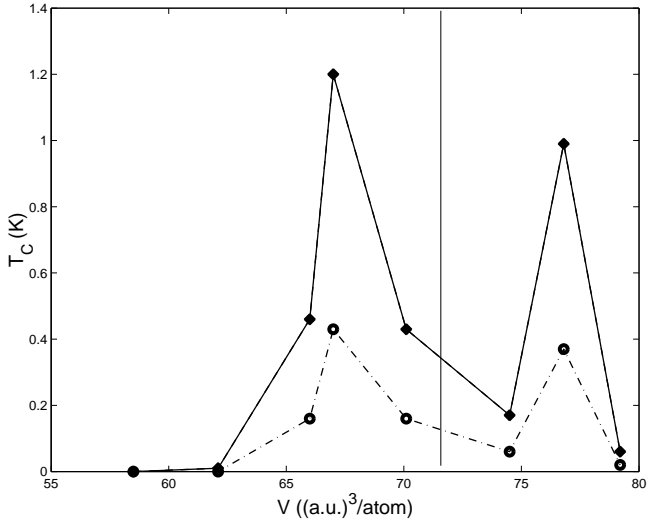}\end{center}
\caption{
 Calculated $T_C$ in hcp iron from eq. \ref{eq:tcsp} using the values $\lambda_{ep}$ and
 the weighted values of $\lambda_{sf}$ and $\omega_{log}$ in Table 1. The full
 line is for $\lambda_{sf}^0=0$ and the broken line for $\lambda_{sf}^0=\lambda_{sf}^1$.
 To the right of the vertical line is the stable structure bcc (and
 T$_C$=0) according to the calculations. 
Superconductivity  has been observed with $T_C$ below 2 K in a pressure range
of about 150 kBar starting near the bcc-hcp transition pressure \cite{shim}.
 }
\end{figure}

\begin{figure}[tb!]
\leavevmode\begin{center}\epsfxsize8.6cm\epsfbox{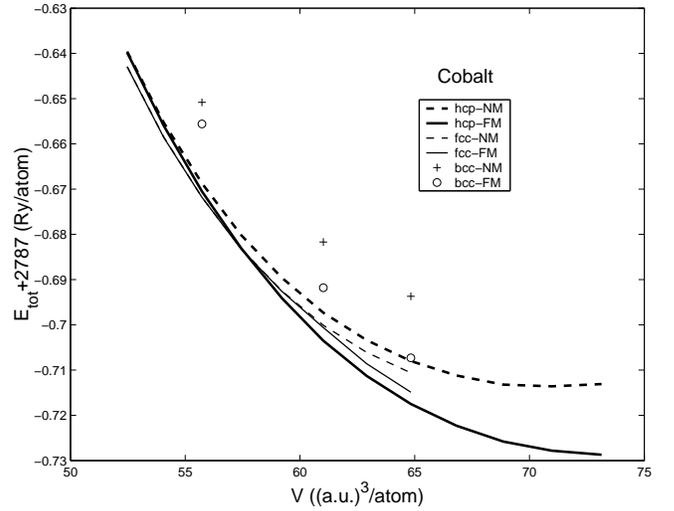}\end{center}
\caption{
 Calculated total energies using GGA potential as function of the volume for
 hcp, fcc and bcc Co.
 }
\end{figure}

\begin{figure}[tb!]
\leavevmode\begin{center}\epsfxsize8.6cm\epsfbox{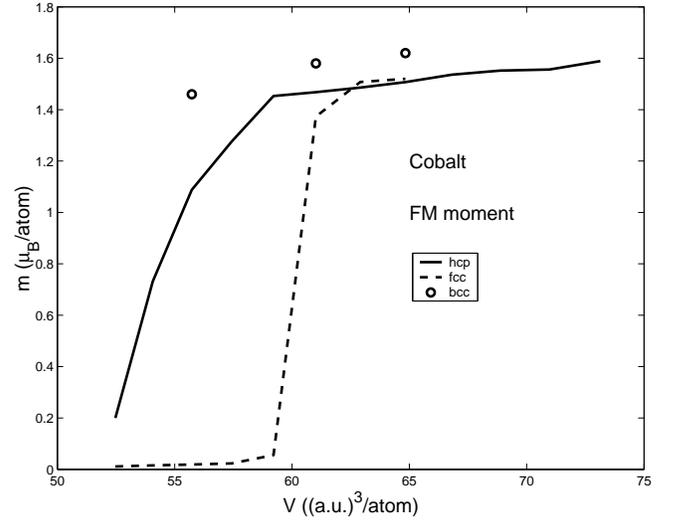}\end{center}
\caption{Magnetic moments calculated using GGA potential for hcp, fcc and bcc Co.
 }
\end{figure}

\end{document}